\documentclass[sigconf, screen]{acmart}
\usepackage{booktabs,stfloats,microtype} 
\usepackage{enumitem}

\usepackage{comment}
\usepackage{multirow}
\usepackage{array}
\usepackage{float}
\usepackage{stfloats}
\usepackage{booktabs}  
\AtBeginDocument{%
  \providecommand\BibTeX{{%
    \normalfont B\kern-0.5em{\scshape i\kern-0.25em b}\kern-0.8em\TeX}}}

\copyrightyear{2024}
\acmYear{2024}
\setcopyright{rightsretained}
\acmConference[DIS '24]{Designing Interactive Systems Conference}{July 1--5, 2024}{IT University of Copenhagen, Denmark}
\acmBooktitle{Designing Interactive Systems Conference (DIS '24), July 1--5, 2024, IT University of Copenhagen, Denmark}\acmDOI{10.1145/3643834.3661576}
\acmISBN{979-8-4007-0583-0/24/07}

\usepackage[autostyle,german=guillemets]{csquotes}

\newcommand{\revision}[1]{{\color{black}\textbf{}#1}}
\newcommand{\eg}[0]{\textit{e.g., }}
\newcommand{\ie}[0]{\textit{i.e., }}
\makeatletter
\usepackage{hyperref,quoting}
\quotingsetup{vskip=0pt,font={itshape,raggedright},rightmargin=0pt}

\begin{document}

%\title{How to Convey AI Explanations}
%\title{Developing the Answer Bank: How AI Explanations can be designed to address end-user needs}
% \title[The Development of \textit{Answer Bank}]{The Development of \textit{Answer Bank}: How AI Explanations Can Be Designed to Address End-User Needs}
  %\title[]{The AI-DEC: Designing Explainable AI}
  %\title[]{AI-DEC: User-centered design tool for context-tailored XAI}
  \title[The AI-DEC: A Card-based Design Method for User-centered AI Explanations]{The AI-DEC: A Card-based Design Method for\\User-centered AI Explanations}

%Supporting structural design prototyping for context-tailored XAI
%Supporting structural user explanation designs for context-tailored XAI
%Supporting structural explanation design for context-tailored XAI


\author{Christine P Lee}
\orcid{0000-0003-0991-8072}
\affiliation{%
  \institution{Department of Computer Sciences University of Wisconsin--Madison}
  \country{Madison, Wisconsin, USA}
}
\email{cplee5@cs.wisc.edu}

\author{Min Kyung Lee}
\authornote{Equal senior contribution.}
\orcid{0000-0002-2696-6546}
\affiliation{%
  \institution{School of Information\\The University of Texas at Austin}
  \country{Austin, Texas, USA}
}
\email{minkyung.lee@austin.utexas.edu}

\author{Bilge Mutlu}
\authornotemark[1]
\orcid{0000-0002-9456-1495}
\affiliation{%
  \institution{Department of Computer Sciences University of Wisconsin--Madison}
  \country{Madison, Wisconsin, USA}
}
\email{bilge@cs.wisc.edu}
%%\renewcommand{\shortauthors}{Trovato and Tobin, et al.}

\begin{abstract}
Increasing evidence suggests that many deployed AI systems do not sufficiently support end-user interaction and information needs. Engaging end-users in the design of these systems can reveal user needs and expectations, yet effective ways of engaging end-users in the AI explanation design remain under-explored. To address this gap, we developed a design method, called \textit{AI-DEC}, that defines four dimensions of AI explanations that are critical for the integration of AI systems---communication content, modality, frequency, and direction---and offers design examples for end-users to design AI explanations that meet their needs. We evaluated this method through co-design sessions with workers in healthcare, finance, and management industries who regularly use AI systems in their daily work. Findings indicate that the AI-DEC effectively supported workers in designing explanations that accommodated diverse levels of performance and autonomy needs, which varied depending on the AI system's workplace role and worker values. \revision{We discuss the implications of using the AI-DEC for the user-centered design of AI explanations in real-world systems.}

\end{abstract}

\begin{CCSXML}
<ccs2012>
   <concept>
       <concept_id>10003120.10003121.10003122.10003334</concept_id>
       <concept_desc>Human-centered computing~User studies</concept_desc>
       <concept_significance>500</concept_significance>
       </concept>
   <concept>
       <concept_id>10003120.10003121.10011748</concept_id>
       <concept_desc>Human-centered computing~Empirical studies in HCI</concept_desc>
       <concept_significance>500</concept_significance>
       </concept>
   <concept>
       <concept_id>10003120.10003123.10010860.10010859</concept_id>
       <concept_desc>Human-centered computing~User centered design</concept_desc>
       <concept_significance>500</concept_significance>
       </concept>
   <concept>
       <concept_id>10003120.10003123.10010860.10010911</concept_id>
       <concept_desc>Human-centered computing~Participatory design</concept_desc>
       <concept_significance>500</concept_significance>
       </concept>
   <concept>
       <concept_id>10003120.10003123.10010860.10011694</concept_id>
       <concept_desc>Human-centered computing~Interface design prototyping</concept_desc>
       <concept_significance>500</concept_significance>
       </concept>
 </ccs2012>
\end{CCSXML}

\ccsdesc[500]{Human-centered computing~User studies}
\ccsdesc[500]{Human-centered computing~Empirical studies in HCI}
\ccsdesc[500]{Human-centered computing~User centered design}
\ccsdesc[500]{Human-centered computing~Participatory design}
\ccsdesc[500]{Human-centered computing~Interface design prototyping}
%\begin{CCSXML}
%<ccs2012>
%  <concept_id>10010520.10010553.10010562</concept_id>
%  <concept_desc>Computer systems organization~Embedded systems</concept_desc>
%  <concept_significance>500</concept_significance>
%  <concept_id>10010520.10010575.10010755</concept_id>
%  <concept_desc>Computer systems organization~Redundancy</concept_desc>
%  <concept_significance>300</concept_significance>
%  <concept_id>10010520.10010553.10010554</concept_id>
 % <concept_desc>Computer systems organization~Robotics</concept_desc>
%  <concept_significance>100</concept_significance>
%  <concept_id>10003033.10003083.10003095</concept_id>
%  <concept_desc>Networks~Network reliability</concept_desc>
%  <concept_significance>100</concept_significance>
%</ccs2012>
%\end{CCSXML}

%\ccsdesc[500]{Computer systems organization~Embedded systems}

\keywords{Design cards, user-centered design, human-AI interaction}


\begin{teaserfigure}
    \includegraphics[width=\textwidth]{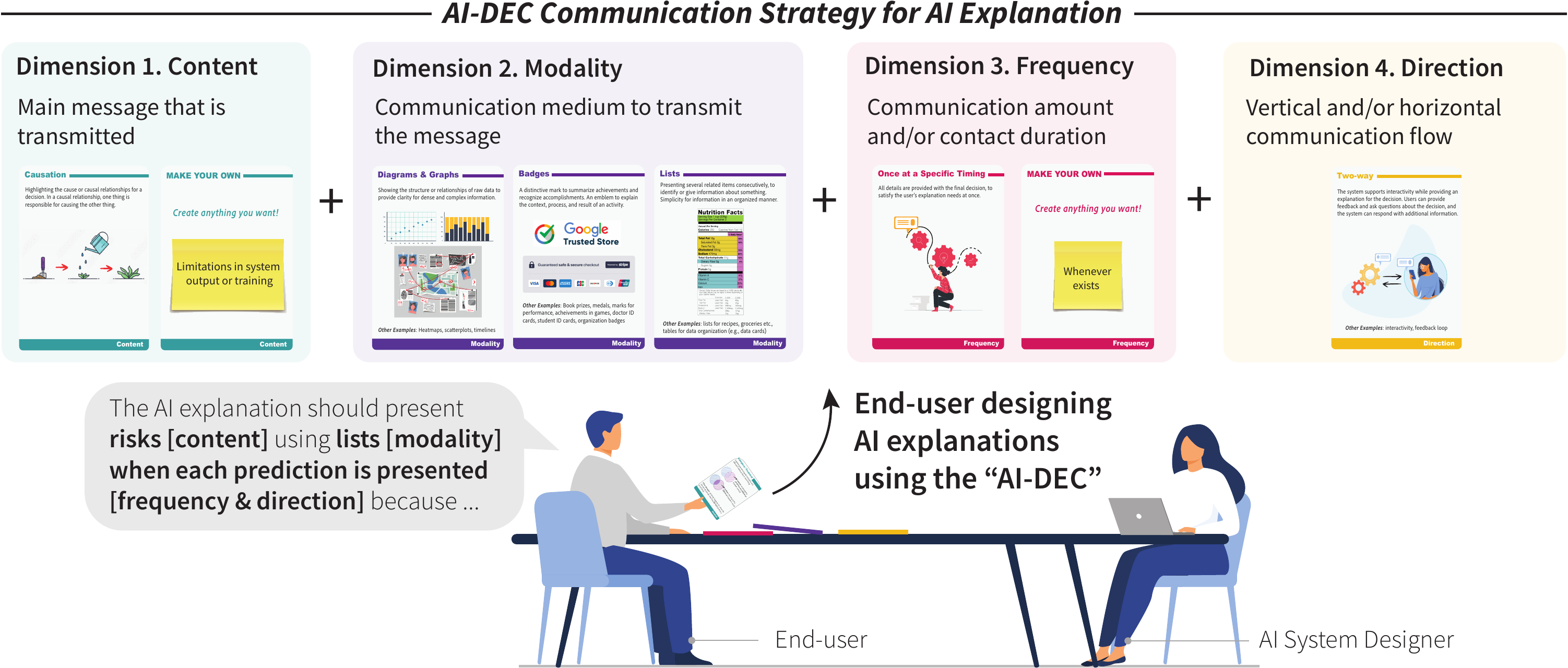}
   \vspace{-12pt}
  \caption{ 
  We introduce \textit{AI-DEC}, a card-based design method for user-centered AI explanations. AI-DEC presents design examples for key communication dimensions of AI explanations: content, modality, frequency, and direction. %This method incorporates end-users in crafting tailored AI explanations, enabling them to develop communication strategies aligned with their needs. 
  The figure provides an example of our co-design session where a worker (\ie end-user) utilizes the AI-DEC to design personalized AI explanations.
  % meeting their information and interaction needs.
  \Description{A user collaborates with an AI system designer at a table, utilizing AI-DEC cards from four dimensions: content, modality, frequency, and direction. They are designing a user-centered AI explanation, arranging the cards to outline the structure of an explanation that presents risks through lists each time a prediction is made.}
}
  \label{fig:teasor}
\end{teaserfigure}

\maketitle

\section{Introduction}

Artificial Intelligence (AI) systems are increasingly deployed into real-world applications across many domains, including healthcare  \cite{yu2018artificial, jiang2017artificial, hu2023investigating}, finance \cite{dhieb2020secure, sheikh2020approach, cirqueira2020scenario}, management \cite{chien2020artificial, nadikattu2021influence}, education \cite{holmes2022state, alam2023harnessing, ho2024s}, and government \cite{ahn2020artificial, kankanhalli2019iot}. Such an increase is largely due to the rapid advancement in recent machine learning (ML) and AI fields, enabling AI systems to provide decision support or final decisions on workers' behalf. However, such deployment is often accompanied by adaptation and integration challenges \cite{weiner2022ai}, due to a mismatch of worker expectations, organizational issues, and technical constraints \cite{westenberger2022failure, schlegel2023failure}. A majority of these constraints and limited success of integration stem from a lack of consideration for end-users---users who actually engage with the AI system. These technologies are often developed without sufficient consideration of the deployed environment, end-users' interaction and information needs, and how users will use and adopt the AI system.

There has been increasing interest in designing AI systems to be human-centric, with a focus on making them comprehensible to end users and meeting their needs \cite{capel2023human, liao2021human}. Several design methods have been used to facilitate user-centered AI, including participatory design \cite{lee2019webuildai}, co-design \cite{lee2022unboxing, zhang2023stakeholder, ayobi2021co, gu2021lessons, liu2023human}, field studies \cite{goodman2021toward, jiang2022needs, kaltenhauser2020you}, and design workshops \cite{andersen2022collaborative, park2022designing, stapleton2022imagining, sun2022investigating, kuo2023understanding, yang2020re}. However, there are still limited methods for effectively engaging end-users in explainable AI (XAI) design. Prior work has emphasized the importance of understanding user needs, situational factors, and deployed environment to design accessible and comprehensible AI explanations\cite{wolf2020designing, ehsan2021expanding, liao2020questioning}; yet little research has engaged end-users in AI explanation design process. This difficulty is partially due to a limit in intuitive and accessible design methods that render AI explanations as accessible design material for end-users. Design methods that facilitate prototyping or collaboration with end-users have the potential to design AI explanations that better support various workplaces.

%\cite{, liao2021human}

%Prior work has emphasized the importance of understanding user needs, situational factors, and deployed environment to design accessible and comprehensible eXplainable AI (XAI) experiences for end-users \cite{wolf2020designing, ehsan2021expanding, liao2020questioning}. 
%However, there are still limited methods for effectively engaging end-users in XAI design. While an active line of research has explored methods to make AI as accessible design materials for designers \cite{yang2020re, liao2021human}, relatively little research has been done to engage end-users in AI design with the exception of \citet{kuo2023understanding, liu2023human}. 

In this work, we developed a card-based design method called the \textbf{\textit{AI-DEC}}, \revision{(\ie \textbf{AI} Explanation \textbf{De}sign \textbf{C}ards)} to specifically meet end-users informational and organizational needs for designing effective AI explanations. The AI-DEC defines four dimensions of AI explanations that are critical for the integration of AI systems in deployment environments: communication content, modality, frequency, and direction. The cards in the AI-DEC provide design examples for end-users to design AI explanations that meet their needs. The aim of the AI-DEC is to directly involve end-users in the design process of AI explanations, supporting them in prototyping and expressing their visions. We then used the AI-DEC for co-design sessions with 16 workers who regularly engage with AI systems in their workplace to explore whether the AI-DEC helped them design AI explanations aligned with their information and interaction needs. We recruited workers from three domains: healthcare, finance, and management to assess the applicability of the AI-DEC in various contexts. Our findings show that the AI-DEC assisted workers in designing AI explanations tailored to their performance and autonomy needs. These needs vary depending on the specific role of the AI system in the workplace and the workers' individual values. Finally, we present implications on how to use the AI-DEC in designing context-adaptive and user-centered AI explanations for real-world applications. Our contributions are the following: 
\begin{itemize}
    \item The AI-DEC, a card-based design method to engage end-users in the design process of AI explanations;
    %\item Incorporation of human communication theory to design human-centered AI explanations;
    \item Using the AI-DEC to co-design AI explanations with workers from various domains who possess empirical experience engaging with AI systems daily;
    \item Insights derived from worker design solutions on information needs for user-centered AI explanations in workplaces.
\end{itemize}

\section{Related Work}

\subsection{Advancements in Human-Centered AI Research and Design Approaches}
An expanding body of research has concentrated on enhancing the human-centered design and interpretability of AI systems. Previous studies have employed diverse methods, such as design probes \cite{wang2022understanding,so2020human}, design workshops \cite{andersen2022collaborative, park2022designing, stapleton2022imagining}, prototyping \cite{mishra2021designing, scurto2021prototyping, subramonyam2021protoai}, Wizard of Oz \cite{viswanathan2022situational}, design workbooks and toolkits \cite{franccoise2021marcelle, biermann2022tool}, participatory design \cite{lee2019webuildai}, co-design \cite{aranda2022co, zhang2023stakeholder, benjamin2022explanation}, research through design \cite{morrison2021social, schroder2021unboxing}, and end-user evaluations \cite{garbett2021towards, lu2021expert, schmalzle2022harnessing, zheng2022telling}, resulting in guidelines for creating human-centered AI systems. Noteworthy contributions include the integration of human decision-making theory into AI explanations \cite{wang2019designing}, emphasizing the consideration of situational factors in AI explanations \cite{ehsan2021expanding}, utilizing uncertainty for transparency \cite{bhatt2021uncertainty}, supporting end-user participation in AI decision-making \cite{lee2019webuildai}, and providing guidelines for communicating AI trustworthiness \cite{liao2022designing}. These contributions collectively advance the goal of making AI systems more interpretable, thereby promoting their effective integration into various domains while addressing concerns regarding transparency, trust, and user engagement.

Simultaneously, another line of research has emerged, focusing on developing design tools and approaches to aid designers in crafting human-centered AI. Educational materials have been devised to enhance designers' understanding of AI system capabilities and technical intricacies. Interactive machine learning tools facilitate designers in comprehending technical capabilities and simulating user-system interactions \cite{van2018prototyping}. Guidelines on addressing failures and managing uncertainty have been proposed. Design methods to comprehend user needs and foster collaboration with users have also been introduced. Examples include a ``Question Bank'' categorizing explanation needs and related user queries \citet{liao2020questioning}, a scenario-based design method addressing context-specific user requirements in complex settings \citet{wolf2019explainability}, and a repository of research results and tools from the DARPA XAI program to aid designers in incorporating XAI techniques \citet{hu2021xaitk}. Additionally, data probes \cite{subramonyam2021towards, wang2022understanding} have been developed to facilitate AI co-design sessions with end-users. While work on understanding the needs for the user-centered design of AI systems has been actively conducted, there are limited design tools and methods developed to directly incorporate and support end-users in the design process of human-centered AI systems.

\subsection{Card-based Design Methods in HCI}

Design cards have been recognized as valuable approaches in HCI and design research \cite{nadal2022tac, hsieh2023cards, hanington2019universal}. \citet{roy2019card} categorized the main purpose of 155 sets of design cards into systematic design methods, human-centered design, or domain-specific design. Notable examples include IDEO Method Cards \cite{ideoMethod, ideoNature}, Envisioning Cards from the Value Sensitive Design Lab \cite{friedman2012envisioning}, and the Design with Intent toolkit \cite{lockton2010design,lockton2013design}. The appeal of design cards lies in their simplicity, tangibility, and ease of use, making them an accessible tool to infuse information and inspiration into the design process \cite{lucero2016designing, wolfel2013method, bekker2011developmentally, hornecker2010creative}. The main advantages of using design cards involve deepening design knowledge, fostering creativity, engaging stakeholders, and refining design practices \cite{roy2019card, hornecker2010creative,carneiro2012ilo, mueller2014supporting}. A thorough literature review of design cards by \citet{hsieh2023cards} identified seven different types of design knowledge that design cards are used to present: Creative Inspiration, Human Insights, Material \& Domain, Methods \& Tooling, Problem Definition, Team Building, and Values in Practice. However, the authors also highlight the need to overcome perceived barriers of designers that design cards lack usefulness and alignment with practical requirements. To optimize the effectiveness of design cards, there is a call for developing cards with usability and applicability in mind, incorporating clear guidelines, concrete examples, well-defined keywords, and flexibility for designer customization. Moreover, efforts should focus on supporting design stages beyond ideation, specifically in prototyping, implementation, and evaluation. Employing the advantages of intuitive operation, versatility, and tangible interaction, the AI-DEC utilizes design cards as a method to enhance end-user usability and applicability across various contexts.

\begin{table*}[!htbp]
    \caption{\textit{AI-DEC Design Elements ---}
    Description of design elements and related works in HCI, XAI, ML, and design that develop, use, or evaluate these elements.
    In each dimension, the ``Design your own'' card allows users to create their own design elements.
    }
    \Description{A comprehensive table detailing design elements, descriptions, and their origins related to the AI-DEC tool. The elements are organized by content, modality, frequency, and direction, each offering a 'Design your own' option for personalized element creation.}
    \label{design_elements}
    \centering \small
    \renewcommand{\arraystretch}{1.4}  
    \begin{tabular}{>{\raggedright\arraybackslash}m{0.15\textwidth} >{\raggedright\arraybackslash}m{0.6\textwidth} >{\raggedright\arraybackslash}m{0.15\textwidth}}
         \toprule
    \textbf{Design Element} & \textbf{Description} & \textbf{Related Works} \\
         \toprule
         \multicolumn{3}{c}{\textbf{Dimension: Content}} \\
         \toprule
         System Metrics & Measures model precision, quantifies feature influence, assesses stability, and analyzes model complexity & \cite{lundberg2017unified, datta2016algorithmic, lim2010toolkit} \\
         \midrule
         Endorsement & Validates information through peer and expert approval, using reliable sources and established knowledge & \cite{henry2022human, lee2019webuildai, yang2023harnessing} \\
         \midrule
         Counterfactual & Shows how altering variables or hypothetical changes impacts model outcomes & \cite{byrne2019counterfactuals, miller2019explanation, warren2022features, chou2022counterfactuals} \\
         \midrule
         Causation & Identifies and clarifies the causal links or decision-making rules between input features and model outputs & \cite{beckers2022causal, warren2022features, chou2022counterfactuals, ghorbani2019towards, vilone2021quantitative} \\
         \midrule
         Subgoal \& Breakdown & Dissects complex models into manageable subgoals, individual components, or hierarchical relationships & \cite{das2023subgoal, ahn2022can} \\
         \midrule
         Compare \& Contrast & Compares similar and differing cases to highlight decision-making patterns, anomalies, and critical factors & \cite{zhang2022towards, labaien2020contrastive, robeer2018contrastive, mittelstadt2019explaining} \\
         \midrule
         Examples & Uses specific instances and use cases to illustrate model operations & \cite{van2021evaluating, keane2019case, kenny2021explaining} \\
         \toprule
         \multicolumn{3}{c}{\textbf{Dimension: Modality}} \\
         \toprule
         Metaphor & Utilizes familiar concepts to simplify and clarify complex ideas & \cite{khadpe2020conceptual,murray2022metaphors, dhanorkar2021needs, pierce2018addressing, hey2008analogies} \\
         \midrule
         Badges & Represents status, achievements, affiliations, skills, or recognition & \cite{zichermann2010game, kidwell2016badges} \\
         \midrule
         Lists & Organizes information into structured, consecutive formats & \cite{kelley2009nutrition, li2022understanding, emami2021informative} \\
         \midrule
         Abstraction \& Summary & Condenses complex model information into more manageable forms, often excluding intricate details & \cite{choi2023concept, ribeiro2018anchors, kim2018interpretability} \\
         \midrule
         Natural Language & Uses everyday language to communicate complex concepts & \cite{ahn2022can, yang2023harnessing, lakkaraju2016interpretable, brown2020language} \\
         \midrule
         Icon \& Pictures & Provides visual representations of model elements (\eg data, results, and concepts) & \cite{zikmund2014blocks, zikmund2010demonstration, price2007communicating, okan2015improving} \\
         \midrule
         Diagrams \& Graphs & Visually depicts model structures, decision processes, variable relationships, trends, and patterns & \cite{brown2011health, barnes2016tailoring, okan2012higher} \\
         \toprule
         \multicolumn{3}{c}{\textbf{Dimension: Frequency}} \\
         \toprule
         Once at a Specific Timing & Delivers at strategically chosen moments (\eg before/after predictions, periodically, triggered by events, during critical decisions or educational needs) & \cite{ribeiro2016model, ribeiro2016should} \\
         \midrule
         Progressive Disclosure & Gradually reveals information to manage complexity and cognitive load by strategic needs or user prompts & \cite{springer2019progressive, springer2020progressive} \\
         \midrule
         On-demand & Provides information responsive to users' requests & \cite{bussone2015role, wallis1984customized} \\
         \toprule
         \multicolumn{3}{c}{\textbf{Dimension: Direction}} \\
         \toprule
         One-way & Supports unidirectional information delivery from the source to the recipient, without necessitating any interaction or response & \cite{ribeiro2016model, ribeiro2016should} \\
         \midrule
         Two-way & Supports interactive communication that allows recipients to engage, inquire, and participate actively & \cite{bertrand2023selective, ouyang2022training} \\
         \midrule
         Multi-way & Supports a dynamic interactive environment and information exchange not only within the system but also among multiple users & \cite{ehsan2021expanding, schoonderwoerd2021human} \\
         \bottomrule
    \end{tabular}
\end{table*}

\section{Development of the AI-DEC}

In this section, we discuss the development procedure of the AI-DEC. Section \ref{sec:structure} explains how the structure of the AI-DEC was built from a theoretical model of communication strategy. Subsequently, Section \ref{sec:elements} discusses the procedure of selecting the design elements of the AI-DEC.% driven from HCI, XAI, ML, and design literature.  

\subsection{Communication Model as a Design Framework for AI Explanations} \label{sec:structure}
%ai explanation needs to be designed with a communication principle in mind. 
%why human-human communication principles can be effective for ai explanation design

\revision{In settings where humans interact with one another, communication acts as the binding element that holds together various distributed channels \cite{mohr1990communication, mohr1994characteristics}. Its crucial role lies in facilitating the smooth flow of information, coordination, and decision-making among the different entities involved. Ineffective communication can lead to problems within these channels, causing individuals to feel excluded or frustrated due to misunderstandings \cite{mohr1995communication}.}
The importance of communication extends to the deployment of AI systems in real-world environments where they engage with end-users \cite{hu2022polite}. The interaction between the AI system and end-users encompasses complex policies, numerous factors, and training processes along a pathway that is often not communicated to end-users due to the opaque nature of AI systems.
Therefore, it is essential to design AI explanations with a communication principle in mind. Crafting AI explanations that center around communication strategies tailored to the deployment environment and end-user needs can enhance understanding, adoption, and satisfaction among end-users. This approach can also help address skepticism and distrust that might hinder the successful integration of AI systems during deployment, promoting transparency, effective functionality, and the seamless integration of the AI system.

The AI-DEC's structure is based on the theoretical model of communication strategies proposed by \citet{mohr1990communication} for organization and marketing channels. 
This model for channel communication reflects the diversity of channel settings and elicits the role of communication to attain enhanced levels of channel outcomes. 
In this context, a channel refers to the ``channel of distribution,'' which is the pathway or route through which goods or services move from the producer or manufacturer to the end consumer. It encompasses the various intermediaries, such as wholesalers, retailers, and other middlemen, who facilitate the movement of products along the distribution process. 
The authors propose a contingency theory, in which communication strategy moderates the impact of channel conditions (structure, climate, and power) on channel outcomes (coordination, satisfaction, commitment, and performance). When a communication strategy matches the channel conditions, channel outcomes will be enhanced in comparison with the outcomes when a communication strategy mismatches channel conditions. 
Specifically, the communication strategy consists of a combination of communication dimensions of context, modality, frequency, and direction \cite{farace1977communicating}.  \textit{Content} describes the ``main message that is transmitted --- or what is said.'' \textit{Modality} represents the ``medium of communication or the method used to transmit the information.'' \textit{Frequency} involves the ``amount of communication and/or duration of contact between organizational members.'' Finally, \textit{direction} refers to the ``vertical and/or horizontal movement of communication within the organizational hierarchy.''
\begin{figure*}[!h]
  \includegraphics[width=\textwidth]{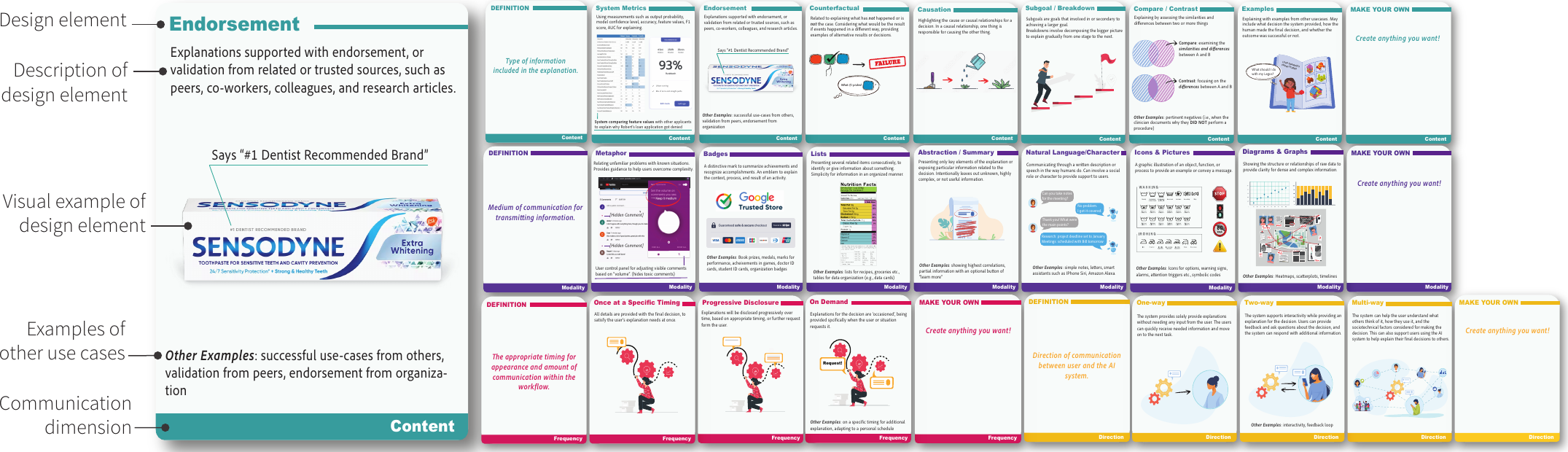} 
  \caption{\textit{Example of AI-DEC Design Card ---} 
  \textit{Left:} The figure depicts a sample design card in the AI-DEC. The card is color-coded to represent its category within the communication dimensions. It includes a comprehensive description of the design element, a visual example, and examples of other use cases. \textit{Right:} The complete AI-DEC set is displayed. For further details, please see Figure \ref{fig:cardCollection} in the Appendix.
  }
  \Description{An illustration showing two aspects of the AI-DEC design tool. On the left is a detailed view of a single card from the content dimension named 'endorsement', which includes a description, visual example, and other use cases. On the right, a complete set of AI-DEC cards is displayed.}
  \label{fig:cardDetail}
%   \vspace{-6pt}
\end{figure*}

\subsection{Design of the AI-DEC} \label{sec:elements}

\revision{From the communication model of \citet{mohr1990communication}, we adopt its four-dimension structure, including content, modality, frequency, and direction, along with their associated definitions, to create a card-based design tool for AI explanation design. The AI-DEC incorporates this structured approach as a card deck, where each dimension consists of cards containing specific design elements.}

%The AI-DEC consists of four types of cards, content, modality, frequency, and direction in line with the model proposed by \citet{mohr1990communication}. Each facet of the AI-DEC (\ie content, modality, frequency, and direction), contains cards with design elements. 

%The selection of the design elements for each card was derived from existing design research in ML, HCI, human-robot interaction (HRI), and XAI. We conducted a review of existing research to incorporate used design features, findings, and implications into the design cards. 

%To develop the design cards, we conducted a focused literature review on user-centered design, human-computer/AI interaction, interpretable ML, and explainable AI. We utilized major academic databases such as ACM Digital Library, Google Scholar, IEEE Xplore, and Springer, selecting papers from the past two decades. The search employed keywords such as ``user-centered design'', ``transparency in AI'', ``human-AI collaboration'', and ``explainable AI''. Initial screening was based on titles and abstracts for relevance, followed by full-text reviews to verify their suitability for inclusion as design elements. This review was designed to be comprehensive but not exhaustive, aiming to gather a robust set of design elements that can inspire end-users imaginations.
%//our goal is not an extensive review of all design elements!
\revision{To develop the design cards, we conducted a focused review on user-centered design in human-computer/AI interaction, interpretable ML, explainable AI, and design literature. This review was designed to be comprehensive but not exhaustive, aiming to gather a robust set of design elements that can inspire end-users imaginations. Table \ref{design_elements} lists the design elements incorporated into the design cards for each communication dimension, detailing the descriptions of each element and the literature sources from which they were derived. Based on the literature chosen, the design cards encompassed a variety of explanation techniques, including interpretation methods applied to machine learning model outputs or computational systems, human communication and reasoning strategies, data visualization techniques, and more.}
As the selection of the design elements is not exhaustive, each dimension included a ``Design Your Own!'' card where users can freely create their design element. Figure \ref{fig:cardDetail} shows an example of a design card. The complete set of the AI-DEC can be found in Figure \ref{fig:cardCollection} in the Appendix. 

%The full list, specific description, and the past use in related works of the design elements included in the AI-DEC are shown in Table \ref{design elements}. 
%While a portion of the AI-DEC is shown in Figure \ref{expCards}, the full set of design cards can be found in the Appendix. 

The objective of the AI-DEC is to be utilized in design studies, incorporating end-users into the process of crafting user-centric AI explanations and fostering structured, creative thinking. According to the classification of card-based design tools by \citet{roy2019card}, the AI-DEC falls under ``human-centered design,'' where the cards help ``designers to focus on the users of a product, service, or system, considering their needs, wishes, and requirements.''

When using the AI-DEC, end-users choose cards from the four communication dimensions to construct their desired explanation from the AI system. The final explanation design comprises a combination of cards from each dimension, representing a design solution tailored to user information or interaction needs. This design solution encompasses the explanation's content, mode of delivery, timing or frequency of presentation, and necessary interactivity or user intervention. An example of a user's AI explanation design using the AI-DEC is shown in Figure \ref{fig:cardToTheme}.
\revision{A printable copy of the full AI-DEC is accessible in the supplementary materials.\footnote{The supplementary materials can be found at \url{https://osf.io/fqpy2/?view_only=97dcc454d9dd455c94dd03178c984bbe}}}

%\section{Evaluation Method}
\section{Co-designing AI Explanations with the AI-DEC}
\revision{
To understand the effectiveness and usability of the AI-DEC, we conducted co-design sessions with end-users of AI systems to create AI explanations tailored to their specific needs and contexts. 
We chose a co-design approach to demonstrate examples of how the AI-DEC can be used in real-world applications that incorporate AI systems. Furthermore, this approach aligns with the AI-DEC's goal of facilitating collaborative and inclusive design.
It actively engages participants in the design process while providing valuable insights into user interactions and perceptions of the AI-DEC, helping us understand user engagement and identify areas for improvement.
}
\subsection{Study Procedure} 

%Through the co-design sessions, we aimed to understand how the AI-DEC can be used to support users in designing AI explanations that align with their information and interaction needs. Moreover, we wanted to understand how useful the AI-DEC was when used across different use contexts, where user needs may differ. 
%We conducted a co-design study where participants created AI explanations using the AI-DEC. 
\revision{
The co-design study comprised two parts: understanding \textit{what} workers wanted to design for, and observing their use of the AI-DEC to generate design solutions on \textit{how} they wanted AI explanations to be communicated, in order to satisfy their information and interaction needs. Initially, to understand what to design for, participants reviewed the ``Question Bank'' by \citet{liao2020questioning}. This bank includes a taxonomy of user needs for AI systems, which we used to help workers easily identify their explanation needs. Utilizing the Question Bank, workers identified five explanation categories that represented their most crucial explanation needs, based on their interactions with the AI system in their workplace.}

Based on their selection, participants then developed AI explanations using the AI-DEC. Participants freely selected cards from each dimension (\ie content, modality, frequency, and direction) as many as they wanted, or created their own design element using the ``Design your own!'' card. Once participants finished their explanation design, they were asked to describe their design solution in detail and envision how it would fit into their workflow. After the design session, participants answered interview questions to evaluate the AI-DEC. The study lasted approximately one hour. 

%\christine{During our co-design session, participants first ranked their explanation needs, derived from their experience of interacting with AI systems everyday which identified contextual needs for AI explanations from where the system falls short. The contextual needs varied depending on the participants' workplace environment, workflow, and individual requirements. To address the contextual needs, participants used the AI-DEC to prototype explanation designs that provided an improved and ideal response from the AI system. }

\subsection{Participants}
%We recruited workers who actively used AI systems in their daily work, as their insights and experience established a valuable baseline for designing AI explanations that supported users' specific needs.
We recruited knowledge workers who interact with decision-support AI systems as an integral part of their daily workflow. To understand the applicability and usefulness of the AI-DEC across various contexts, 16 participants were recruited from three different work domains: health (6), finance (5), and management (5). The three domains were chosen as they are among the most utilized areas where decision-support AI systems are applied in real-world workplaces \cite{chui2018notes,chui2017artificial}. Participants were recruited through university mailing lists and study flyers posted in workplaces in the United States. A pre-screening survey determined eligibility based on workers' domain, workplace tasks of interacting with a decision-support AI system, and period of interaction experience with a minimum of three months. Each study session took place in person with one participant and one researcher.
Participants were provided with a \$50 USD payment upon completion of the study. Participants (9 male, 7 female) were aged 21--47 ($M=31.1$, $SD=8.2$). Demographic data is included in Table \ref{participant information}. In presenting our findings, we denote participant work domain with ``H'' (H1--H6) for healthcare, ``M'' (M1--M5) for management, and ``F'' (F1--F5) for finance.

\begin{table*}[!h]%[!t]
    \caption{\textit{Participant Information ---}
    Demographics and background information.
    }
    \Description{A table listing participant details for a study on AI-DEC usage. Information includes participant ID, age, gender, race, job title, and the AI system used in their workplace.}
    \label{participant information}
    \centering
    \renewcommand{\arraystretch}{1.3}
    \small
    \begin{tabular}{m{0.05\textwidth}m{0.05\textwidth}m{0.1\textwidth}m{0.1\textwidth}m{0.26\textwidth}m{0.26\textwidth}}
         \toprule
   \textbf{ID}& \textbf{Age}& \textbf{Gender}& \textbf{Race}& \textbf{Job}& \textbf{System Use} \\
         \midrule
        \multicolumn{6}{c}{\textbf{Domain: Health}} \\
        \midrule
 
 \textbf{H1}   & 34 & Female & White  & Radiologist & Computer Aided Detection\\
  \hline
  \textbf{H2}   & 26 & Female & White  & Acute Care Physician & Decision Support Healthcare Software\\
  \hline
  \textbf{H3}   & 26 & Female & White  & Emergency Room Physician & Decision Support Healthcare Software\\
  \hline
  \textbf{H4}   & 30 & Female & White  & Primary Care Physician & Decision Support Healthcare Software\\
  \hline
  \textbf{H5}   & 27 & Male & White  & Family Medicine Physician & Decision Support Healthcare Software\\
  \hline
  \textbf{H6}   & 46 & Male & White  & Radiologist & Computer Aided Detection\\
  \midrule
    \multicolumn{6}{c}{\textbf{Domain: Management}} \\
        \midrule
  \textbf{M1}   & 20 & Female & Hispanic  & Multinational Chain Crew Member & Automated Scheduling System\\
  \hline
  \textbf{M2}   & 23 & Male & White  & Multinational Chain Crew Member & Automated Scheduling System\\
  \hline
  \textbf{M3}   & 30 & Male & Hispanic  & Human Resources Recruiting Director & Applicant Tracking System\\
  \hline
  \textbf{M4}   & 41 & Female & White  & Marketing Manager & Advertisement Management Software\\
  \hline
  \textbf{M5}   & 37 & Male & White  & Supply Chain Analyst & Automated Scheduling System\\
  \midrule
  \multicolumn{6}{c}{\textbf{Domain: Finance}} \\
        \midrule
  \textbf{F1}   & 25 & Male & Hispanic  & Banker & Automated underwriting system\\
  \hline
  \textbf{F2}   & 35 & Male & White  & Loan Officer & Automated underwriting system\\
  \hline
  \textbf{F3}   & 47 & Male & White  & Banker, Branch Manager & Automated underwriting system\\
  \hline
  \textbf{F4}   & 31 & Male & White  & Banker, Branch Manager & Automated underwriting system\\
  \hline
  \textbf{F5}   & 21 & Female & Black  & Banker & Automated underwriting system\\
 \bottomrule
    \end{tabular}
    % \vspace{-12pt}
\end{table*}

\subsection{Analysis}
All user studies were held in person and video recorded using Zoom. Session recordings were then transcribed using the Otter.ai automated speech-to-text generation tool and manually checked by the research team. The transcriptions were coded following the guidelines developed by \citet{clarke2014thematic} and \citet{McDonald19}. 
We conducted Thematic Analysis (TA) using the transcriptions and participants' design solutions using the AI-DEC\footnote{The initial interview phase was analyzed in a separate study. This paper focuses on the co-design sessions of the research.}. The first author was familiarized with the data by facilitating the studies and initially generating codes. Through iterative discussions with the whole research team, categories of grouped codes were generated and refined. The categories were then further clustered and reflected upon with the entire research team to derive themes emerging from our study data. After all candidate themes were discussed and reviewed, the final themes are reported as findings.

\begin{figure}[!h]
  \centering
  \includegraphics[width=\columnwidth]{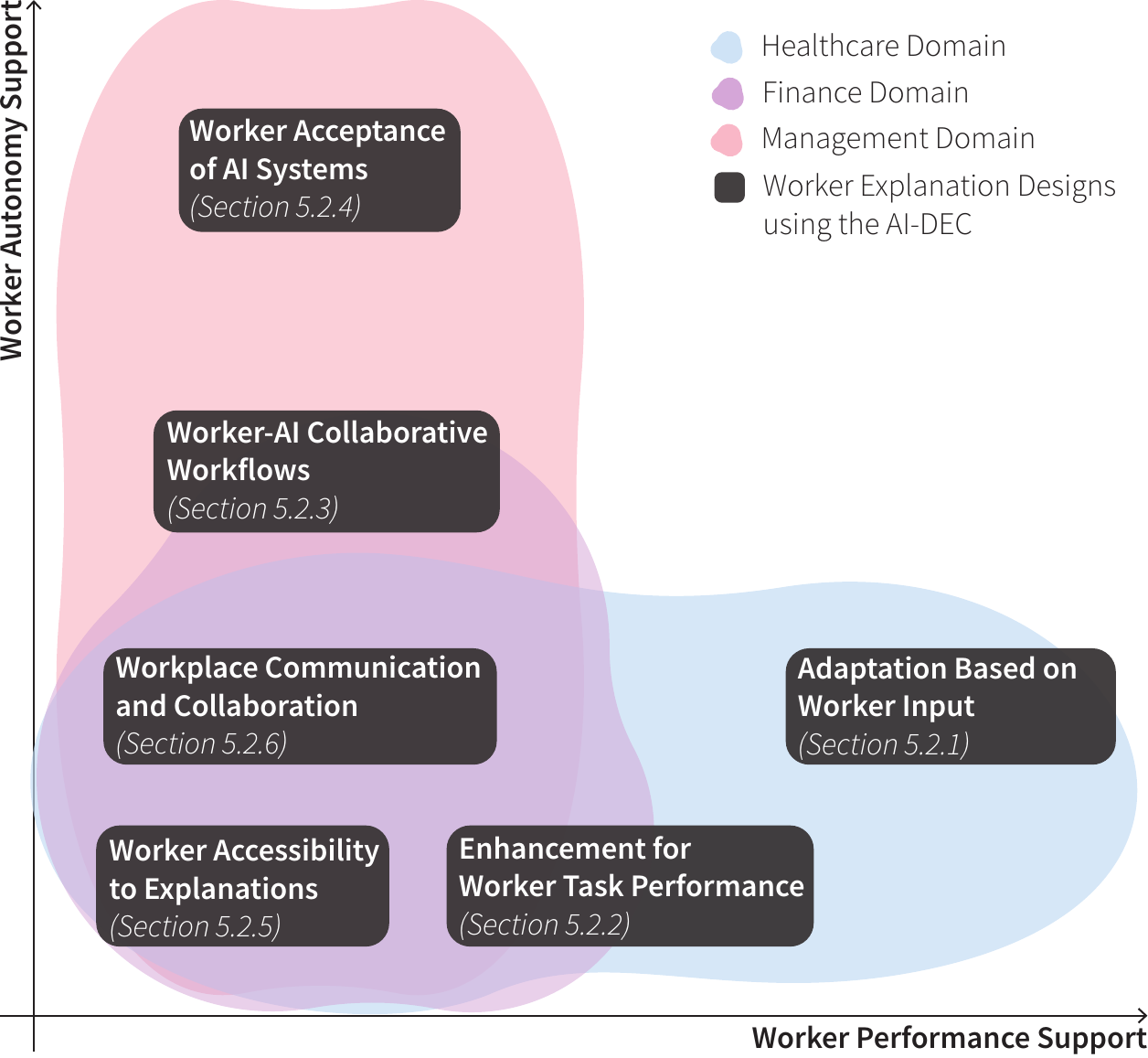} 
  \caption{\textit{Findings Summary ---} The graph shows different goals of the explanations that workers designed using the AI-DEC. The goals varied in the level of support that workers desired for their autonomy and performance across work domains. Healthcare workers focused explanation designs on enhancing their performance. The finance workers aimed to design explanations that improved both their performance and autonomy. In management, workers prioritized explanations that supported their autonomy.
}
\Description{Graph summarizing the goals of AI explanations designed using AI-DEC across different domains. The x-axis represents support for worker performance and the y-axis for worker autonomy. Different abstract shapes indicate how goals vary across sectors like healthcare, management, and finance.}

  \label{fig:findings}
\end{figure}
%The graph shows different goals of the explanations that workers designed using the AI-DEC. The goals varied in the level of support that workers desired for their autonomy and performance across work domains. 

%Workers' explanation design goals differed depending on the role of the workplace AI system and the worker's needs. 
\section{Findings}

In this section, we present the findings of the co-design session. 
In Section \ref{sec:role}, we describe how the workers' explanation needs differed across domains depending on the AI system's role in the workplace and worker values. 
In Section \ref{sec:designs}, we present the AI explanation designs created by workers using AI-DEC. Depending on the unique information and interaction needs, these designs aimed to support various levels of worker performance and autonomy.
Finally, Section \ref{sec:evaluation} describes the workers' evaluations of the usefulness, shortcomings, and potential improvements for future use of the AI-DEC.

\subsection{Workplace AI Systems and their Role} \label{sec:role}
In this section, we provide an overview of the AI systems participants use in their everyday work settings. This understanding provides essential context for understanding the participants' design motivation and the findings. These workplace-incorporated systems leverage AI techniques, such as machine learning, data analysis, and reasoning algorithms. The primary objective of these systems is to provide decisions or decision-making support to workers. Despite the workplace AI systems having similar functionalities of decision-support, they had different roles and responsibilities in each work domain. Furthermore, the different roles of the AI system led to workers feeling different levels of autonomy. Factors including workplace characteristics, worker's autonomy afforded by the AI system's role, and worker values shaped diverse information and interaction needs among workers. 
%The details of these findings are presented in a different publication. The differences in workers' interaction experiences with their workplace AI systems led to workers selecting different categories from the question bank for explanation support. Depending on their needs, the categories spanned to support worker autonomy or worker performance. 

\paragraph{Healthcare}

Workers in the healthcare domain use Clinical Decision Support Systems (CDSS) and AI systems that support the digitization, management, and sharing of patient medical records while streamlining clinical workflows. The system also provides clinical decision-support tools, offering workers alerts, reminders, and evidence-based guidelines for lab tests, treatments, and diagnosis at the point of decision-making. The computer-aided diagnosis (CAD) systems help workers interpret medical images, particularly in detecting and analyzing potential abnormalities, patterns, or lesions, by leveraging rule-based algorithms and ML techniques.

The AI system in the healthcare domain had an assistive role, leaving the final decision-making to workers. Workers also had follow-up tasks using the output of the AI system in addition to making the final decision, which involved explaining their final decision to various stakeholders in the workplace. %Workers did not express apprehension about diminishing opportunities to apply their expertise; instead, they envisioned the AI system playing an assistive role in enhancing worker performance in tasks like decision support, action planning, and routine automation within their workflow.

\paragraph{Finance}

Workers in the finance domain use Automated Underwriting Systems (AUS) to analyze loan applications based on financial information (\eg credit reports, digital transaction records, prior loan history, income, etc.) and generate decisions on approving or denying mortgages, personal loans, small business loans, auto loans, and student loans. During customer interactions, workers gather financial information from the customer, enter it into the system, and receive a definitive decision from the system with no or limited warnings for risks (\eg ``multiple high-risk factors'').

The role of the AI system in the finance domain was more decisive than in healthcare, as it held the authority to approve customer applications. This shift in decision-making power led workers to perceive a reduction in autonomy, limiting their ability to apply expertise as they did before the introduction of AI systems. However, workers continued to fulfill essential duties such as explaining the AI system's outcome to customers and advising financial plans. %Consequently, there was a desire among workers for the AI system to enhance support for their performance in fulfilling these responsibilities.

\paragraph{Management}

Workers from multinational chains use an automated system to schedule shifts and positions for various tasks. The AI system automatically generates shift schedules based on factors like performance, skill set, and generated profit.
Workers in manufacturing companies use centralized software programs to automatically manage resources and scheduling, including allocating the required labor type, labor time, and tasks to workers while providing information about available resources and task processing.
An advertising manager uses a social media management platform that monitors and analyzes social media to recommend the best times for individuals and businesses to promote advertisements and manage multiple accounts.
A worker in a Human Resources department uses automation software to manage and streamline the recruitment and hiring process, particularly for candidate screening to analyze resumes and assess qualifications.

The AI systems in the management domain played the most authoritative role compared to the healthcare and finance domain, as it made the final decision for workers regarding decision outcomes, worker tasks, and worker schedules. Workers had minimal or no opportunities to independently make decisions or apply their expertise, resulting in a low level of autonomy provided to workers. %This limited autonomy led workers to express the need to address the limited autonomy imposed by the AI system.

\begin{figure*}[!h]
  \includegraphics[width=\textwidth]{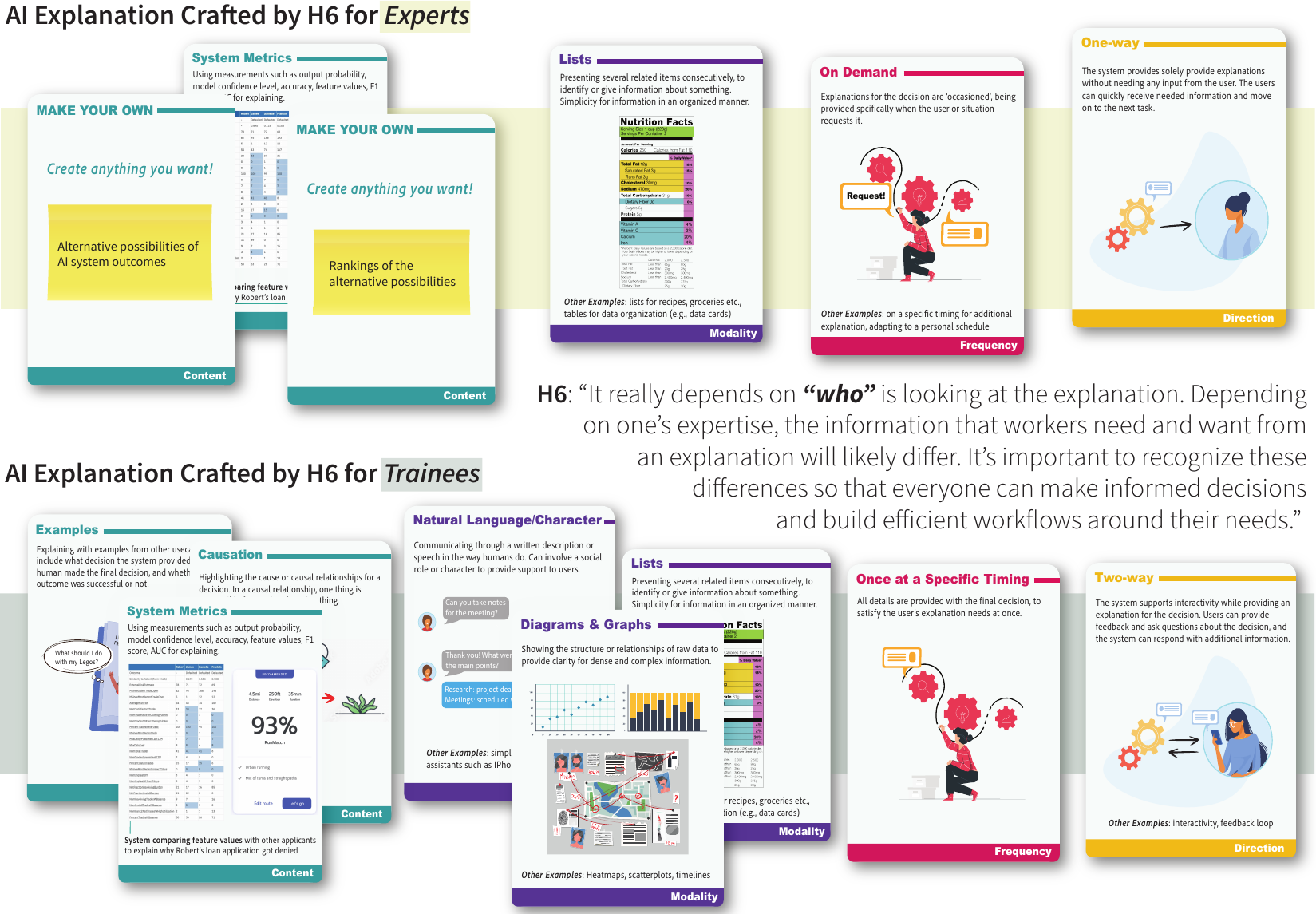} 
  \caption{\textit{Worker's AI Explanation Design Using the AI-DEC ---} 
  The figure shows an AI explanation crafted by a worker (H6). The worker developed two distinct types of explanations for the worker's varying levels of expertise. \textit{Top:} for experts, the AI explanation functioned as an extra eye. \textit{Bottom:} for trainees, it served as an educational tool, providing more detailed information. Details can be found in Section \ref{sec:exptypes}.
  }
  \Description{Comparison of two AI explanation designs created using AI-DEC. The top design, by experts, features metrics, alternative outcomes, and on-demand lists. The bottom design, intended for trainees, includes natural language explanations, diagrams, and interactive elements scheduled for specific times.}

  \label{fig:cardToTheme}
%   \vspace{-6pt}
\end{figure*}

\subsection{Workers' AI Explanation Designs with the AI-DEC}\label{sec:designs}
%AI Explanations that Workers Designed Using the AI-DEC

In this section, we report workers' explanation designs using the AI-DEC. From our analysis, we identified six themes in the workers' explanation designs, which diverged across different levels of supporting worker performance and autonomy.
In this work, we refer to performance as the effectiveness, efficiency, and quality of work carried out by workers in their domains, while autonomy refers to the degree of independence, self-governance, and decision-making authority granted to workers within their roles.

Using these definitions, we identified six explanation themes from workers' designs aimed at enhancing performance, autonomy, or both. The explanation design themes include:  (1) adaptation based on worker input; (2) enhancement for worker task performance; (3) worker-AI collaborative workflows; (4) worker acceptance of AI systems; (5) worker accessibility to explanations; and (6) workplace communication and collaboration. A summary of the findings is shown in Figure \ref{fig:findings}. The complete set of workers' explanation designs can be found in Figure \ref{fig:participantDesign2} in the Appendix.

%Specifically, in the healthcare domain, explanation designs were tailored to enhance worker performance. In the finance domain, explanation designs aimed to improve performance while also supporting worker autonomy. In the management domain, explanations were designed to prioritize supporting worker autonomy.

\subsubsection{Adaptation Based on Worker Input}
%Adaptable Explanations by Worker Input
In the healthcare domain, workers developed AI explanations that allowed for worker input in response to the AI system's output to support worker performance. %Workers identified shortcomings in existing workplace AI systems, noting their limited provision for user input or feedback in response to the AI system's output. This lack of interactivity rendered the AI system less effective in supporting worker performance and led to worker skepticism regarding its utility for their tasks.

\paragraph{Input to Improve Performance of AI System}

Using the AI-DEC, healthcare workers (H1, H3, H5, H6) designed AI explanations to improve the system's decision-making performance by incorporating their feedback, particularly regarding error corrections and adjustments to decision-making features.
\revision{For example, one worker (H3) described \textit{``So the goal here is to prevent me from mistakes. So with the output, there needs to be reasoning for my sensemaking to determine if it [AI system] is correct, we use graphs and metrics, but I also want a breakdown...[selecting cards] But most importantly there needs to be iterative feedback, so I want two-way (direction), and I want a list of what was fixed.''} They emphasized that the AI system's output accuracy was directly linked to their own task performance, leading to select design features that enabled continuous improvement based on error correction from worker feedback. }

Additionally, workers (H5, H6) sought validation of whether the AI system adhered to values and practices similar to those of human workers in decision-making processes. For example, one worker (H5) used the AI-DEC to devise explanations that conveyed system limitations and risks associated with error correction and worker awareness.
The explanation design comprised causation and limitations (content), diagrams and badges (modality), occurring whenever necessary (frequency), and supporting two-way communication for interactive feedback (direction). In response to the limitations outlined in the explanations, workers envisioned correcting misaligned features to mitigate potential adverse outcomes and increase workers' awareness of risks associated with the AI system's outputs.

\paragraph{Input for Expertise-based Explanations}\label{sec:exptypes}

%Workers additionally designed explanations to facilitate worker input in adjusting factors related to their workflow, aiming to enhance worker efficiency in their performance.
%First, workers developed explanations to enable worker input in managing task allocation between themselves and the AI system. Workers identified repetitive, minor tasks in their routine, such as administrative duties, which could be streamlined through automation by the AI system. The explanation by one worker (H2) involved breakdowns of worker routines to identify areas suitable for AI automation (content), a list for possible tasks (modality), on-demand access (frequency), and bidirectional communication (direction).

Workers (H1, H3, H6) also used the AI-DEC to design explanations that provided different information support based on the worker's level of expertise. \revision{For example, H6 explained \textit{``My design needed to involve multiple versions, because even within my team, explanation needs and uses are different depending on one's expertise. It really depends on ``who'' is looking at the explanation.''} For experts, the explanation aimed to serve as an extra eye, whereas for trainees, it functioned as an educational tool, offering more detailed information.} 
Consequently, one worker (H6) developed two types of explanations using the AI-DEC, when explaining the rationale behind the AI system's outputs. For experts, the worker selected alternative possibilities, metrics, and rankings (content), presented in a list format (modality), on-demand (frequency), and provided one-way communication (direction). Designs focused on ruling out other possibilities and providing concise support for accuracy. Conversely, for novices, the worker selected causation, metrics, and examples (content), presented through lists, diagrams, and natural language (modality), delivered at specific timings at the end of each AI system output (frequency), and facilitating two-way communication for further questions (direction.) The design focused on supporting detailed investigation on the factors and procedures of the final decision. The importance of tailored explanations based on the worker's expertise level was emphasized to ensure that all workers could learn and make informed decisions.

\subsubsection{Enhancement for Worker Task Performance}
Workers in the healthcare and finance domains designed AI explanations to improve worker performance by helping them plan and execute follow-up tasks using the output of the AI system. %They described that explanations provided by the current workplace AI systems lacked clarity in providing support for workers to explain the final decisions to various stakeholders and were insufficient in offering insights for subsequent action planning. This ambiguity often led workers to feel uncertain and dissatisfied regarding the quality of the service they provided in these tasks.

\paragraph{Stakeholder Explanations}
%Workers designed explanations to provide information to use when explaining the worker's final decision (healthcare domain) or the AI system's final decision (finance domain) to various stakeholders.
Workers (H2, H3, H4, F2, F3, F4) used the AI-DEC to create explanations that helped them clearly explain the final decision to various stakeholders in the workplace.
The design elements varied depending on the intended audience, specifically whether explaining to domain experts like co-workers or to domain novices such as customers and patients.
For example, one worker (H3) used the AI-DEC to create two distinct versions of explanations depending on whether the audience was domain experts or novices. 
For domain experts, more technical language involving metrics, counterfactuals, and examples (content) was selected to be presented through intuitive lists and diagrams (modality), be available on demand (frequency), and be presented one-way (direction) as a ``report'' of the worker's decision-making procedure. Such explanations aimed to convey factual details that support the worker's final decision and factors of alternative options regarding ``why not,'' for transparent communication and preventing misunderstandings between workers. 

Conversely, for novices, designs were aimed to easily support high-level understanding. The design entailed causation, breakdown, and comparison (content) for abstracted but informative descriptions, conveyed via metaphors and diagrams (modality) for easier understanding, to be available on demand (frequency) in need of support, and in one-way communication (direction). Explanations for novices clarified outcome rationales while simplifying complex details with interpretable design elements.

\paragraph{Action Planning}
Additionally, the AI-DEC was used to devise explanations to support workers (H1--H4, H6, F1--F5) in strategizing follow-up actions based on the AI system's outputs. They noted that a crucial aspect of their job often involved determining the next steps using the AI system's output to accomplish workplace tasks or objectives (\eg gaining customer approval or planning patient treatment). Thus, workers crafted AI explanations that offer practical guidance, including suggestions on actions required to achieve alternative outcomes. For example, one worker (H1) designed \textit{``recommendations for the ``safest'' next steps based on prior experience or probability,''} while another worker (F2) designed \textit{``test functions'' to experiment with different features for varied outcomes.''} This worker's (F2) explanation design enabled workers to experiment with alternative inputs to observe the impact on outcomes. Selected design cards included causation and counterfactuals (content), lists (modality), available on demand (frequency), and facilitating interactive two-way communication (direction). The design rationale was explained as, \revision{\textit{F2: ``So testing with counterfactuals enables me to ask ``What if I made this change? Can I increase my loan amount? Would I still be in the parameters for the inspection waiver?'' Along with compare and contrast, show me the difference. This gives me the full picture I need to move forward.''}} 

Another worker in the healthcare domain designed explanations to clearly understand all possible outcomes, aiding decision-making in their high-stakes environment. They mentioned, \textit{H6: ``Knowing the possible outcomes, rankings or probabilities becomes much more useful in my workplace where mistakes can be bad. So I selected a list of possible outcomes. Can you give me those as a ranking, or probabilities of each one, and show me the distributions of those where the patient falls? To help me be more sure.''} Accordingly, they used the ``design your own!'' card to include a list of possible outcomes ranked by probability.

\subsubsection{Worker-AI Collaborative Workflows}\label{sec:workerintervention}
Workers in the finance and management domains designed AI explanations to improve worker autonomy when engaging with workplace AI systems. Designed explanations focused on promoting cooperative opportunities between workers and the AI system. %Workers explained that the absence of opportunities to independently make decisions or participate in decision-making processes had a negative effect on their autonomy and made them feel undervalued. Thus, workers utilized explanations to create opportunities for workers to engage in decision-making and leverage their skills with the AI system, thus mitigating the limited autonomy provided by AI systems.

\paragraph{Opportunities for Worker Expertise}

Workers (F2, F3, F5, M2--M5) used the AI-DEC to design explanations that fostered collaboration between AI systems and workers, aiming to augment worker autonomy by providing chances for workers to exercise their expertise. One worker (M1) commented, \revision{\textit{``There are many nuances and contextual factors to consider in daily jobs, where human experience is important. So I chose not to let the AI system handle everything [in my design,] and included a human touch, whether it's [AI explanation] asking what to do or how it did, or actually sharing some responsibilities.'' }}
Thus workers selected design cards to support worker understanding and provide opportunities for worker intervention, specifically envisioned for where the AI explanation sought worker expertise assistance when encountering unforeseen or abnormal situations (\eg bankruptcies, natural disasters, and construction projects). One worker (M2) designed explanations to prompt worker intervention when particularly the AI system faced uncertainty. This explanation design encompassed causation and breakdown (content), summary and list (modality), when encountering uncertainty (frequency), and enabled interactive two-way communication (direction). This explanation aimed to provide information on the tasks completed by the AI system and the obstacles encountered, offering context for workers to intervene and make informed decisions.

\subsubsection{Worker Acceptance of AI Systems}
Workers in the management domain designed explanations to support worker acceptance, aiming to address the lack of autonomy provided by the current workplace AI systems. Such acceptance was an important factor in preventing workers from being reluctant to adhere to the managerial roles and task allocations performed by the AI system.
%Workers described that limited autonomy often led to worker reluctance to embrace the AI system or diminished their inclination to engage with it. Such reactions were due to the lack of consideration of workers' perceptions of usefulness, compatibility with existing practices, and trust in the AI system during its introduction into the workplace. As a result, workers designed explanations to foster rapport and receptivity towards the AI system.

\paragraph{Support for Onboarding of AI Systems}
Using the AI-DEC, workers (M1, M2, M3, M5) crafted explanations that varied in design throughout the deployment timeline. \revision{One worker noted, \textit{M3: ``Explanations need to change depending on how accustomed we are to the system, and how much we know or trust it. Otherwise, it might be too little or too much information.''}} Workers expressed their desire for the design of explanations to evolve based on their growing familiarity and rapport with the system.
For example, one worker (M3) chose ``progressive disclosure (direction)'' to be applied not in a single explanation, but across different stages of the AI system's deployment. They designed different explanation versions for both the introduction and active-use stages.
During the initial onboarding process, the explanation was designed to focus on establishing rapport and fostering workers' understanding of the AI system's capabilities and value. 
Thus, explanation designs encompassed causation of the AI system's output and examples demonstrating the AI system's capabilities in similar use cases along with counterfactuals it considered to show its capabilities (content), conveyed through both summary and extensive lists for detailed understanding (modality), enabling two-way interactive communication to address worker questions for concerns or confusions (direction) any time required on demand (frequency). 
However, during the active use stage, the explanation design entailed more simplified design elements, involving metrics and causation (content), and just summaries or graphs for intuitive understanding (modality), but also supporting user information needs interactively (direction) when needed (frequency). 
Through different designs, the worker envisioned the explanation to gradually unveil less detailed information over time during the deployment, as rapport was built through the introductory explanations.

\subsubsection{Worker Accessibility to Explanations}
In all domains, workers created explanations that included design elements that were readily understandable to laypeople. Workers emphasized the significance of accessibility in explanation design to help workers comprehend and effectively apply it to their workplace tasks. %Workers noted that relying solely on system metrics and performance data did not offer useful or practical information, especially to laypeople. As a result, workers built explanations with design elements that were easily accessible and aligned with specific workflow characteristics.

\paragraph{Accessible Design Elements}
\revision{Workers (H1, H3--H6, F2--F5, M1, M2, M4) selected design cards that focused on easing the understanding and improving the relatedness to workers' daily work experiences to enhance their comprehension and usability of the AI explanation. \revision{For example, one worker in the finance domain (F2) described the need for such explanations as \textit{``My job is to educate borrowers and advise them on how to move forward, but I can't do that if the system doesn't give me meaningful stuff. That's why my explanation gives me specific, financial information that is useful to me, measures that I use every day to do my job.''} This worker's explanation design involved combining the metrics (content) and list (modality) card to include the Debt to Income (DTI) ratio and credit scores, with a list of saving strategies and other loan options.} 

Another worker (M2) from the management domain designed their explanations to reflect the importance of community among workers, emphasizing firsthand experiences and information from colleagues. The design featured interaction examples from employees at various chain locations and used metaphors to articulate the perceived benefits and implications of integrating the AI system. The explanation provided endorsements through examples from others using the AI system under similar circumstances (content), utilize metaphors for relativeness (modality), be available on demand (frequency), through one-way communication (direction). 

One worker (H1) in the healthcare domain designed explanations that aligned with their unique, image-focused workflow practices to help their understanding. They utilized the ``design your own!'' card for color-coding to represent the AI system's uncertainty level, due to its intuitiveness and faster comprehension compared to text. Their explanation design included metrics, examples, and comparisons (content), diagrams and color-coding to support effective worker performance (modality), available on demand (frequency), and facilitated two-way interactive communication (direction). 

}

\subsubsection{Workplace Communication and Collaboration}

Workers across all domains developed explanations to facilitate workplace communication and collaboration. 
Workers emphasized the importance and everyday occurrence of collaborative and communicative interactions in the workplace, as workers regularly cooperated through different responsibilities to achieve shared workspace goals. Moreover, workers also socially interacted and communicated with each other daily to discuss workers' final decisions, offer mentorship, gain insights, or validation. %Worked described the challenges stemming from the existing AI system's inadequate support for these needs, and designed explanations to support the collaborative and communicative work environments.  

\paragraph{Support for Workplace Collaboration}
Workers (H3--H5, F3, F4, M3, M5) designed explanations to enhance collaboration and achieve workplace goals. These explanations focused on monitoring progress, identifying bottlenecks in the pipeline, and adjusting task plans accordingly. For instance, one worker (M5) selected cards that described others' progress (content), included icons within the work chain and summaries of issues (modality), were provided at specific times, such as when obstacles occurred (frequency), and supported multi-way communication (direction). These shared explanations aimed to streamline collaboration and offer guidelines for addressing conflicts. H2 elaborated on such explanation needs as \textit{``Currently, finding where a bottleneck might be is difficult because we don't know when, where, why, and who. It's unclear if someone ignored information from the system, if it malfunctioned, or if another factor was at play. However, if the AI explanation could pinpoint who took what actions and identify any concerns or issues they had, we can act and address them. This teamwork is crucial because it impacts the outcome on patients.''}

Additionally, to further enhance collaboration, healthcare workers (H1, H3, H5, H6) incorporated elements into the explanations that promoted reliability and consensus towards the AI system. Specifically, workers opted for endorsements (content) and badges (modality), noting that societal consensus was crucial in fostering trust in their workplace. Such design elements ensured that all team members could trust and effectively use the system, thereby reinforcing collaborative efforts and improving operational efficiency.

\paragraph{Support for Workplace Communication}
Workers (H1, H3--H5, F1, F5, M1, M2, M4) designed AI explanations to facilitate communication among workers, enhancing social learning and mentoring. These explanations incorporated results from others' interactions and diverse opinions to validate decisions and build trust.
Specifically, one worker (F5), who had recently started their position, designed explanations to incorporate information from past interactions between other workers and the AI system, particularly from higher in the hierarchy and more experienced workers in the workplace hierarchy. They explained their design purpose as \textit{F5: ``It is often uncomfortable to ask my manager every time I want to check something, at the same time I am concerned if I am doing anything wrong, especially if I am learning. So I want the explanation to do it for me, help me learn, and validate as I go so I can develop my skills.''  }
The explanation design included counterfactual options that have been used by others in cases with similar circumstances, a breakdown of how to reach the goal, and examples of decisions made by higher hierarchy workers (content), conveyed through natural language (modality), provided on demand (frequency), and enabling multi-way communication (direction). Workers described that explanations supporting social communication helped facilitate continuous learning and maintain social connections, which were key to building a supportive and innovative work environment.

%\textit{F1: ``Interrogating uncertainty would help me understand what I can do to help the customer best; I can't gain that knowledge if it doesn't let me ask anything.''} %To support the formulation of workers' next steps and plans, categories selected from the Question Bank included Action (4) and How to Still be This (3).
% %\textit{F2: so having a test function with the counterfactuals. What if I made this change? What kind of result do I get? Can I increase my loan amount? What would that do? Would I still be in the parameters for the inspection waiver? Along with compare and contrast, show me the difference.} 
% %\textit{H6: a multi-parametric output including a suite of possible outcomes, rankings or probabilities becomes much more useful. Especially given my workplace where mistakes can be bad. So I selected list, for a suite of possible outcomes. And can you give me those as a ranking? Or probabilities of each one, and show me the distributions of those where the patient falls? To help be be more sure.} 

\subsection{Worker Feedback of the AI-DEC}\label{sec:evaluation}
In this section, we present the workers' assessment of the AI-DEC gathered at the end of the co-design session. 
%, informed by their co-design experience. %Workers articulated the AI-DEC's advantages in fostering user-centered designs, identified shortcomings, and suggested potential improvements and applications for future design sessions.
\subsubsection{Advantages of the AI-DEC for User-centered AI Design}

Workers (H1--H3, H6, M1--M5, F1, F2, F4, F5) noted that AI-DEC effectively translated abstract preferences into practical design solutions. Its four-dimension structure provided constructive guidance, allowing for a systematic design process that addressed a wide range of information needs. One worker explained this perspective as \textit{H3: ``The four categories guided my thinking step-by-step, making me analyze details that I might have overlooked had I simply been asked about my preferences for explanations.''}

\revision{Additionally, the tangible aspects of prototyping the AI explanations were appreciated by workers (H2--H6, M1--M3, M5, F1, F2, F4, F5). Constructing, deconstructing, and restructuring communication strategies facilitated iterative reflection, testing, and adjustment, leading to designs that were practically sufficient and tailored to their unique workplace environments. 

Finally, workers (H1--H3, H6, M2--M5, F2, F5) described the card deck format of AI-DEC as particularly beneficial for its modularity, which supported quick iteration and customization based on feedback and changing requirements. This flexibility not only enhanced the relevance and effectiveness of AI interactions across diverse settings but also encouraged workers to explore creative combinations of design elements, fostering innovative solutions that might not have been possible in more rigid formats.}

%\subsubsection{Improvement Need to Depict AI System's Technical Capabilities}
\subsubsection{Shortcomings and Improvements for the AI-DEC}

Workers (H3, H6, M3, M4) indicated that using the AI-DEC to design worker-centered explanations for unfamiliar AI systems might be difficult due to a lack of understanding of the AI system's capabilities. Those in our study, familiar with their current workplace AI systems, adeptly tailored solutions to meet their needs. However, one worker (M3) noted the potential difficulty in crafting explanations for a new AI system without understanding its capabilities or how it integrates into existing workflows. Workers (H3, H6, M3) suggested that understanding the AI system's capabilities and limitations could be facilitated by providing users with brief lists or summaries of these details before initiating the design process. 

Additionally, workers (H5, M5) observed that certain design descriptions, such as ``abstraction'' and ``examples,'' were excessively wordy and lacked intuitiveness. They proposed utilizing both sides of each card: one side to display additional examples and visual designs for better comprehension and application, and the opposite side to contain detailed descriptions of the design element, offering comprehensive insights. Similarly, other workers (F3, M1) stressed the need to include more examples on each card, showcasing both general and specific use cases relevant to AI systems.

%\subsubsection{Future Use as Probe for Collaboration with AI Designers and Engineers}

For future use, workers (H2, H4, M1, M3, F1) envisioned employing AI-DEC as a tool to collaborate with AI designers and engineers during the design phase of workplace AI systems. They noted their frequent exclusion from the pre-deployment process, despite being the primary users over time. Workers described that the AI-DEC can serve dual purposes through prototyping and sketching tailored AI explanations, as \textit{H5: ``the AI-DEC can act as a middleman to connect expectations and needs with us and the people who develop the AI, it could be a win-win for everyone.''} Workers further elaborated that on the one hand, the AI-DEC can help workers clarify their requirements and desires through hands-on prototyping. On the other hand, it can provide engineers and designers with a clear understanding of user information and interaction needs, fostering effective discussions about the possibilities and limitations of the AI system.

\section{Discussion}

We introduce the AI-DEC, a card-based design method developed to engage end-users and address their needs in the AI explanation design process. Grounded in communication theory encompassing the structure of content, modality, frequency, and direction, the AI-DEC facilitates the development of user-centered communication strategies for AI explanations. Through co-design sessions, we evaluated the effectiveness of the AI-DEC. Our findings suggest that the AI-DEC effectively supported workers in crafting AI explanations suitable for different levels of performance and autonomy needs. These needs varied depending on the AI system's role in their workplace and the values of the workers. Drawing insights from workers' designs and feedback, we present implications for utilizing the AI-DEC to create user-centered AI experiences. 

\subsection{Leveraging the AI-DEC with End-users}

\paragraph{\textit{When} to use the AI-DEC}
\revision{Existing research has shown that XAI techniques can often fall short for end-users requiring domain-specific insights \cite{zhang2020effect, alqaraawi2020evaluating, ehsan2023human}, leading to investigations into the contextual dynamics of deployment such as the AI lifecycle \cite{dhanorkar2021needs, liao2020questioning, liao2021human} and user informational or situational needs \cite{arrieta2020explainable, ehsan2020human, ehsan2021explainable}. A user-centered approach, prioritizing an understanding of users and their environmental characteristics, has been identified as crucial for effective AI system design \cite{ehsan2023human, zytko2022participatory}. Given these insights, we see an opportunity to leverage the AI-DEC to enhance end-user involvement, particularly in scenarios that require effective communication and collaboration with users.

Firstly, the AI-DEC facilitates communication with users about necessary design features for AI explanations, enabling the identification of specific environmental and individual needs. These discussions highlight the unique aspects of the environment, the specific tasks for which the AI system will be used, and the explanation or interaction needs that support users' real-world objectives. For instance, in our study, workers realized that they value the inclusion of ``why not'' factors in explanations to colleagues and sought similar features in their AI explanations. To facilitate these discussions, external design methods or frameworks such as employing a taxonomy of user needs \cite{arya2019one, liao2020questioning}, storyboarding \cite{truong2006storyboarding} use cases, or task-based discussions \cite{kim2024understanding} can be utilized along with the AI-DEC.

The AI-DEC can also act as an effective collaboration tool during various design stages for AI systems, involving a range of stakeholders. In our study, workers envisioned AI-DEC as a probe to enhance communication and collaboration between AI designers and engineers. These probes were envisioned to create a common understanding of user needs, capabilities, and features, as well as the explanations that the AI system should provide. Physical prototypes produced with the AI-DEC were seen to be particularly useful in identifying misunderstandings and encouraging stakeholders to seek consensus through iterative redesigns. For instance, the ``content'' cards were noted by workers to effectively outline the AI system's technical capabilities, helping them envision optimal ways to use the AI systems to meet their needs. This collaborative approach not only uncovers new design possibilities but also helps avert the risk of deployment failures of AI explanations by integrating user perspectives from the outset.
}
\paragraph{\textit{How} to use the AI-DEC}
%Building on our findings, we see the potential for the AI-DEC not only to craft individual explanations but also to design the broader interaction experience between workers and AI systems. While our study required workers to design for single explanation needs (\ie categories from the Question Bank), we observed them broadening their explanation designs to encompass distinct versions for different interaction points with the AI system. 
%For example, workers designed explanations to support different levels of worker involvement based on the level of uncertainty in the AI system's decision-making. Similarly, based on the user's expertise level, different levels of details in content and modality were disclosed at different times to support the worker's decision-making.  
%Workers also considered how explanations should vary by the AI system's deployment phase (\ie introductory versus active use). 
%Similarly, the AI-DEC can be used by end-users to holistically ideate the interaction experience, detailing the role, responsibilities, and communication strategies the AI system would embody.
\revision{
Building on our findings, we recognize the AI-DEC's potential to shape not only individual explanations but also the broader interaction experience between users and AI systems. Although our study required workers to design explanations for specific needs (\ie categories from the Question Bank \cite{liao2020questioning}), we observed them expanding their designs to create distinct versions for different interaction points. For instance, workers crafted explanations that varied in levels of worker involvement based on the uncertainty detected in the AI system's decision-making. Other explanations tailored the content and modality to match the user's expertise, revealing different levels of detail at various stages to support decision-making. Furthermore, workers considered how explanations should adapt across the AI system's deployment phases (\eg introductory versus active use). Thus, the AI-DEC can be used to help end-users holistically ideate the interaction experience, defining the role, responsibilities, and communication strategies that the AI system should embody.

%in their workflow. For example, workers designed explanations to support instances where they would either intervene or allow the AI system complete autonomy. 

Designers can also extend the design elements of AI-DEC to suit workplace requirements or particular preferences. While retaining the established four-dimension framework of content, modality, frequency, and direction, the elements in the cards can be adjusted to include various aspects that designers find valuable or wish to experiment with in relation to their AI projects. This modification enables designers to tailor the AI-DEC cards according to the needs of users, specific deployment domains, problem scenarios, or desired design features. Such adaptability of the AI-DEC also helps address the limitations that can exist with design cards, where the content or elements of the cards lack relatedness to application contexts \cite{hsieh2023cards}.
Nonetheless, expanding the design cards might introduce trade-offs, such as an increased cognitive load on users when generating explanations. In these instances, designers should balance flexibility and complexity. For example, as complexity increases, designers may simplify the design objectives where feasible and offer guidance and tools \cite[\eg][]{liao2020questioning, wang2019designing} to aid users in navigating the designs effectively.
}
%The design process might induce user cognitive load when constructing explanations. The study doesn't discuss this trade-off where increasing XAI Design Cards can support more use cases but induces higher cognitive load." [R2]

\subsection{Application of the AI-DEC}

\revision{

Previous work highlights the challenges of seamlessly integrating AI systems into real-world applications, particularly due to difficulties involving end-users in AI projects and understanding the diverse needs and expectations of various stakeholders \cite{paleyes2022challenges, campion2020managing, weiner2022ai}. We aim to address these difficulties by using the AI-DEC to involve end-users in the design process of prospective AI systems. In this study, we deployed the AI-DEC in workplaces such as healthcare, finance, and management. While the AI-DEC has proven valuable in these professional settings, we believe its potential extends further, potentially aiding the design of AI explanations in a broader range of applications. This is particularly relevant given the increasing research on deploying AI systems in community-level and non-workplace environments \cite{wan2023community}. For example, consider its application in smart cities \cite{tsai2017towards}, where AI systems manage traffic, distribute resources, and ensure public safety \cite{ullah2020applications}. In such complex systems, stakeholders---from city planners to local residents---have varied informational needs and concerns. City officials might need detailed explanations of traffic flow predictions to plan infrastructure changes, whereas residents could benefit from simpler, more direct explanations about how traffic adjustments affect their daily commutes.
The AI-DEC can empower teams to design these varied types of explanations effectively. It guides designers in identifying which aspects of the outputs from the AI system need emphasis and how to communicate them clearly to different stakeholders. Using the cards, designers can determine the optimal level of technical detail for different audiences, ensuring explanations are accessible yet informative. This tailored approach not only enhances transparency but also fosters trust and engagement, making the integration of AI systems more acceptable and supported by the public.

Additionally, the AI-DEC can be utilized to generate knowledge in user experience (UX) and design research. In UX research, designers can conduct surveys and user studies---either in-person or virtual---utilizing AI-DEC to explore users' information needs and encourage them to construct ideal explanations and interaction examples. For virtual studies, a digital version of AI-DEC can allow users to prototype on an online storyboard (\eg Miro \cite{miro2024}). The examples generated, encompassing AI explanations and interactions, can then inform the designs of the final product.

The importance of design research in HCI is increasingly recognized, driven by methodologies such as research through design \cite{zimmerman2007research} and design-focused UX research \cite{vermeeren2016design}. In this context, we believe that the AI-DEC is valuable in generating new knowledge about AI design requirements and in creating artifacts that represent solutions to these challenges. Given the often variable and unclear design requirements or objectives for AI explanations in different deployment settings, the AI-DEC serves as a useful tool in aiding designers to elicit diverse perspectives on human-AI interaction issues, ideate on potential solutions, and engage in reflective iteration to refine these issues and solutions. Importantly, the artifacts produced during these design activities serve as exemplars, offering insights into how future AI explanations and systems should be designed for successful integration. These insights not only enrich our current understanding but also pave the way for future research directions in human-centered AI design.
}

\section{Limitations \& Future Work}
Our study has several limitations to consider. Firstly, the AI-DEC's scope and design features are not exhaustive and \revision{require further validation through comparative evaluations to become a standardized design method. Future work can involve collaboration with design experts to verify and compare AI-DEC against other design tools or methodologies.} Additionally, recruiting a larger number of participants in each domain proved challenging, partly due to the busy schedules of experts and the limited availability of workplaces experienced in using AI systems extensively in their daily operations. Moreover, our work involved three domains utilizing AI systems in real-world applications, excluding other domains such as education and government. Future work should conduct studies with larger pools of participants (\eg both end-users and designers) and a more diverse number of domains to evaluate the efficacy of utilizing the AI-DEC with HCI research methods and the design process of AI systems. Finally, the qualitative and exploratory nature of our research restricts the generalizability of our findings, as they may not apply universally to all user requirements and AI system application contexts.

\section{Conclusion}

In this work, we present the AI-DEC, a card-sorting-based design method developed specifically to meet the informational and organizational needs of end-users for effective AI explanations. The AI-DEC directly incorporates end-users in the design process of AI explanations, allowing them to design AI explanations that fit their personalized needs. We evaluated the AI-DEC through a co-design session with 16 workers from three different domains---healthcare, finance, and management---to understand its efficacy and applicability in designing AI explanations that better support user needs.
Our findings indicate that the AI-DEC helped workers develop explanation designs that supported varying levels of worker performance and autonomy, which differed depending on the AI system's role and worker values. Finally, we present implications for designers using the AI-DEC to build user-centered AI explanations and human-centered AI systems.

\begin{acks}
We thank the anonymous reviewers for their constructive feedback. This research was supported by the National Science Foundation through awards 1925043, IIS-1925043, IIS-1939606, DGE-2125858, CCF-2217721; Good Systems, a UT Austin Grand Challenge for developing responsible AI technologies;\footnote{https://goodsystems.utexas.edu} and Swedish Research Council. Any opinions, findings, conclusions, or recommendations expressed in this material are those of the authors and do not necessarily reflect the views of the National Science Foundation.
\end{acks}
\balance
\bibliographystyle{ACM-Reference-Format}
\bibliography{bibliography}

\appendix
\begin{figure*}[!h]
  \includegraphics[width=\textwidth]{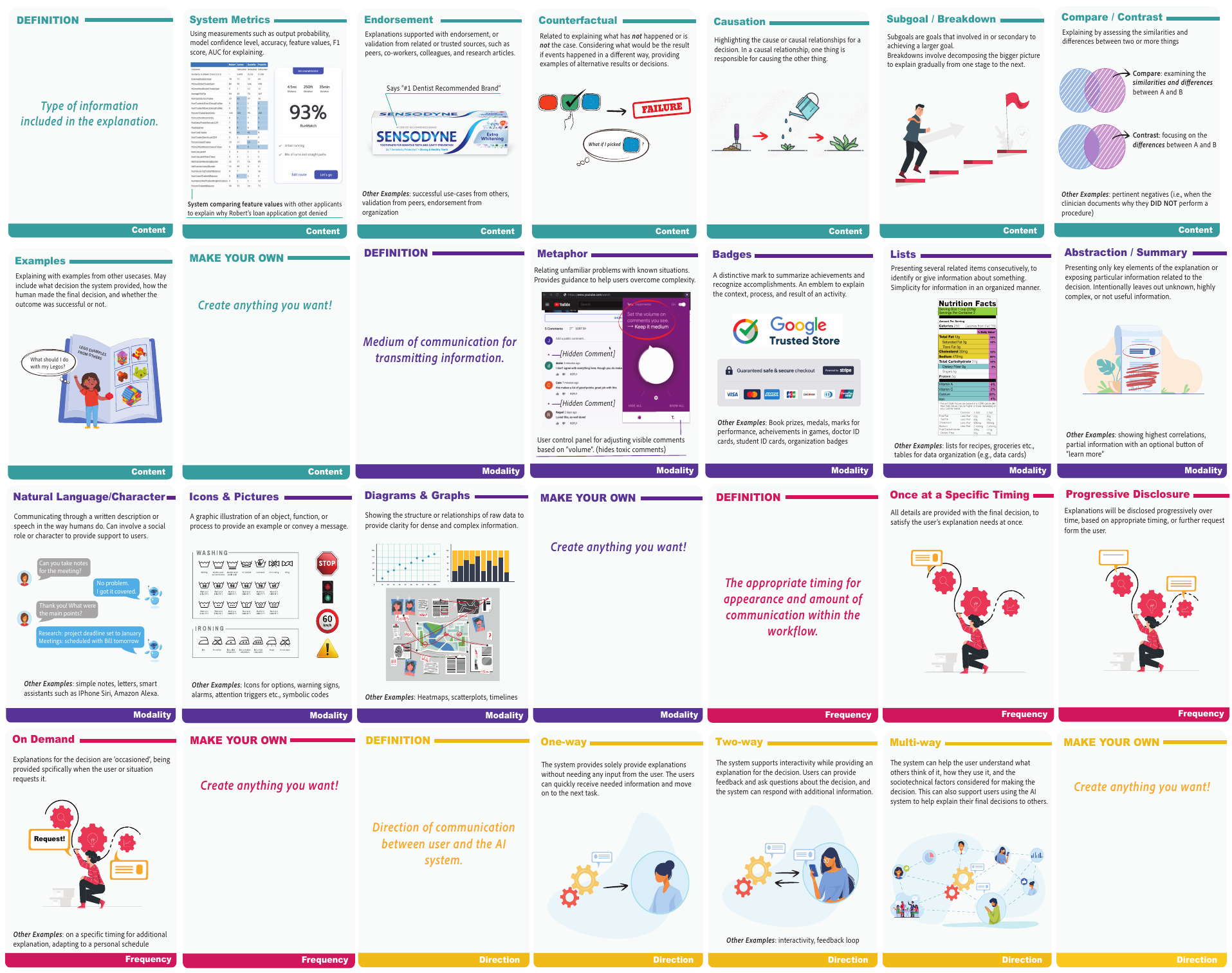} 
  \caption{\textit{Complete Set of AI-DEC ---} 
  The AI-DEC consists of four types of cards, content, modality, frequency, and direction. In each facet, there are cards with design elements derived from existing design research in ML, HCI, human-robot interaction (HRI), and XAI. When utilizing the AI-DEC, end-users will select cards from the four communication facets to assemble their preferred AI explanation. The resulting explanation design will consist of a combination of cards from each facet, forming a tailored design solution to address user information or interaction requirements. This design solution will entail the content of the explanation, the mode of delivery, the timing or frequency of presentation, and the required level of user interactivity. 
  }
  \label{fig:cardCollection}
%   \vspace{-6pt}
\end{figure*}

\begin{figure*}[!h]
  \includegraphics[width=\textwidth]{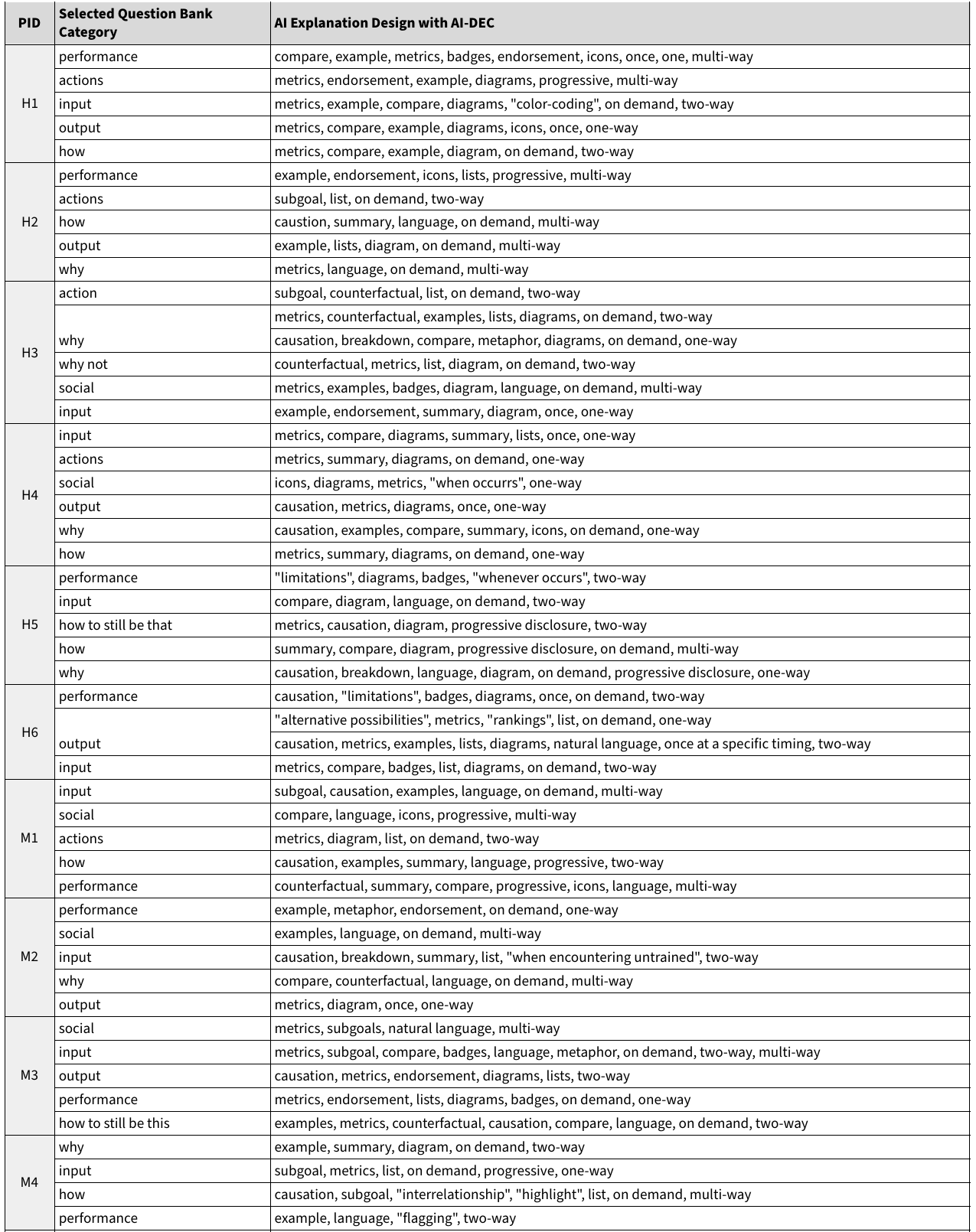}  
 %  \caption{\textit{Full Participant Design Solution ---} 
 % revise
 %  }
  \label{fig:participantDesign}
%   \vspace{-6pt}
\end{figure*}
\begin{figure*}[!h]
   \includegraphics[width=\textwidth]{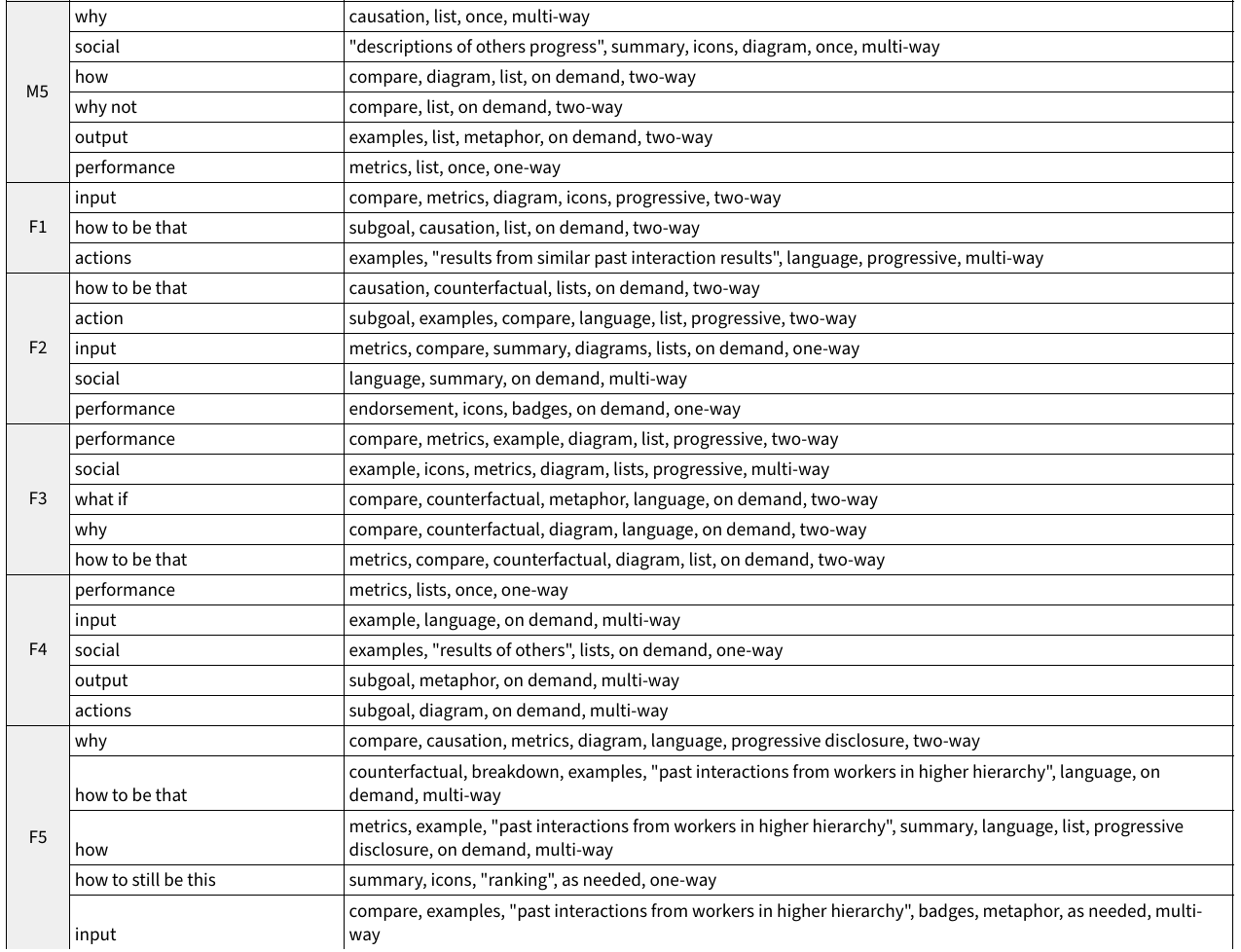} 
  \caption{\textit{Full Set of Participant Design Solution ---} The table presents the complete set of participants' explanation designs utilizing the AI-DEC during the co-design session.
  }
  \label{fig:participantDesign2}
%   \vspace{-6pt}
\end{figure*}

\end{document}